# Annular wave packets at Dirac points and probability oscillation in graphene

Ji Luo, Junqiang Lu, and Daniel Valencia

*Department of Physics, University of Puerto Rico at Mayaguez,*

*Mayaguez, PR 00681, USA*

Wave packets in graphene whose central wave vector is at Dirac points are investigated by numerical calculations. Starting from an initial Gaussian function, these wave packets form into annular peaks that propagate to all directions like ripple-rings on water surface. At the beginning, electronic probability alternates between the central peak and the ripple-rings and transient oscillation occurs at the center. As time increases, the ripple-rings propagate at the fixed Fermi speed, and their widths remain unchanged. The axial symmetry of the energy dispersion leads to the circular symmetry of the wave packets. The fixed speed and widths, however, are attributed to the linearity of the energy dispersion. Interference between states that respectively belong to two branches of the energy dispersion leads to multiple ripple-rings and the probability-density oscillation. In a magnetic field, annular wave packets become confined and no longer propagate to infinity. If the initial Gaussian width differs greatly from the magnetic length, expanding and shrinking ripple-rings form and disappear alternatively in a limited spread, and the wave packet resumes the Gaussian form frequently. The probability thus oscillates persistently between the central peak and the ripple-rings. If the initial Gaussian width is close to the magnetic length, the wave packet retains the Gaussian form and its height and width oscillate with a period determined by the first Landau energy. The wave-packet evolution is determined jointly by the initial state and the magnetic field, through the electronic structure of graphene in a magnetic field.

# I. INTRODUCTION

Graphene is a unique two-dimensional (2D) material where carbon atoms form into hexagons on a plane. Its crystal structure is composed of two 2D hexagonal sublattices and its first Brillouin zone is a turned hexagon in wave-vector space. Near the Dirac points that correspond to the zero-temperature Fermi wave vectors, the energy dispersion is a cone that has two branches connecting at the apex, and an electronic state has two envelope functions that satisfy a Dirac-like equation.[1] The unique atomic and electronic structures of graphene lead to unique properties that have been investigated extensively.[2-5]

Wave packets are fundamental in that their motion determines the electronic transport. For graphene, the zitterbewegung, or the trembling motion of wave packets is of particular interest, as it provides a new material to investigate this phenomenon suggested long time ago and brings about new chances for its experimental observation.[6-8] In a magnetic field, the trembling motion becomes persistent and another interesting feature, the revival of wave packets, appears.[9-12] Furthermore, strain in graphene may generate pseudo-magnetic fields.[13] It was found that for different Dirac points, the pseudo-magnetic fields have different effects on the wave packet states, due to the broken valley symmetry.[14] For experimental observation, femtosecond laser pulses were considered in generating wave-packet states.[15] It was suggested that dipole-moment radiation accompanies the trembling motion, and this may provide a probe to the wave packets.[9,15]

The previous work mainly focused on wave packets whose central wave vectors are not at Dirac points, because trembling motion occurs only when the wave packets have an initial momentum. Formally, these wave packets more resemble quasi-classical particles and the electronic probability is distributed in a localized region. In a magnetic field they demonstrate quasi-classical motion. In this work, wave packets whose central wave vector is at Dirac points are investigated. Different from the localized ones, such wave packets form into annular peaks that propagate to all directions, and the electronic probability is distributed in ripple-like rings. Annular wave packets can be prepared with a magnetic field. Although their average coordinates remain unchanged, they also demonstrate essence of the trembling motion, the oscillation of probability. In graphene, Dirac points correspond to singularities of the energy dispersion where the partial derivatives of the energy with respect to the wave vector do not exist. Hence the annular wave packets cannot be treated by the usual textbook approach in which one expands the energy dispersion into a Taylor series. In this work, direct superposition of electronic eigen-states is performed to compose these wave packets for graphene in free state and in a magnetic field, and numerical calculations are carried out to investigate their propagation. Features of such wave packets are demonstrated by the calculation results, and relations between these features and the electronic structure of graphene are presented.

## II. CONSTRUCTION OF WAVE PACKETS IN GRAPHENE AT DIRAC POINTS

In general, a wave packet in a crystal is the superposition of Bloch functions that have adjacent wave vectors. By expanding the energy dispersion into a Taylor series, one concludes that the wave packet moves like a quasi-classical particle with the velocity $\vec{v}_g(\vec{k}) = \hbar^{-1} \nabla_{\vec{k}} E(\vec{k})$, where $\vec{k}$ is the central wave vector.[16] For graphene, however, the energy dispersion near a Fermi wave vector $\vec{k}_F$ is $E_\pm = \pm \hbar v_F \sqrt{(k_x - k_{F,x})^2 + (k_y - k_{F,y})^2}$, where $v_F = 10^6 \, m/s$ is the Fermi speed. It follows that $\partial E_\pm / \partial k_x$ and $\partial E_\pm / \partial k_y$ do not exist at $\vec{k}_F$. Hence at Dirac points, one cannot expand the energy dispersion into a Taylor series and obtain the quasi-classical motion of the wave packet. Besides, the connection of the two branches of the energy dispersion at $\vec{k}_F$ may further complicate the problem.

For $\vec{k}_F$ as the central wave vector, nevertheless, one may still compose wave-packet states directly by the superposition of states with adjacent wave vectors. Hereafter we use $\vec{k}$ to denote $\vec{k} - \vec{k}_F$. Near Dirac points, the electronic eigen-states of graphene are described by two envelope functions $\psi_1$ and $\psi_2$ that satisfy the Dirac-like equation

$$v_F \begin{pmatrix} 0 & \hat{p}_x \mp i\hat{p}_y \\ \hat{p}_x \pm i\hat{p}_y & 0 \end{pmatrix} \begin{pmatrix} \psi_1 \\ \psi_2 \end{pmatrix} = E \begin{pmatrix} \psi_1 \\ \psi_2 \end{pmatrix}, \qquad (1)$$

where $\hat{p}_x = -i\hbar \partial/\partial x$, $\hat{p}_y = -i\hbar \partial/\partial y$, and the upper and lower signs in $\mp$ and $\pm$ are respectively for two nonequivalent Dirac points $K$ and $K'$.[1] Near each Dirac point, the eigen-energy $E$ has two branches

$$E_\pm(\vec{k}) = \pm \hbar v_F \sqrt{k_x^2 + k_y^2} = \pm \hbar v_F k. \qquad (2)$$

For each wave vector, there are two sets of envelope functions

$$\begin{pmatrix} \psi_1^\pm \\ \psi_2^\pm \end{pmatrix} = \begin{pmatrix} C_\pm \\ \pm(k_x + ik_y)k^{-1}C_\pm \end{pmatrix} \exp(i\vec{k} \cdot \vec{r}) \qquad (3)$$

for $K$ and two sets of envelope functions

$$\begin{pmatrix} \psi_1^\pm \\ \psi_2^\pm \end{pmatrix} = \begin{pmatrix} C_\pm \\ \pm(k_x - ik_y)k^{-1}C_\pm \end{pmatrix} \exp(i\vec{k} \cdot \vec{r}) \qquad (4)$$

for $K'$, where + and - in $\pm$ are respectively for positive energy $E_+$ and negative energy $E_-$, and $C_\pm$ are constants. A wave-packet state also has two envelope functions $\Psi_1(\vec{r},t)$ and $\Psi_2(\vec{r},t)$ that are the superposition of $\psi_1^\pm$ and $\psi_2^\pm$. According to Eqs. (2), (3) and (4) one has

$$\Psi_1(\vec{r},t) = \int_\infty C_+(\vec{k}) \exp(i\vec{k} \cdot \vec{r} - iv_F k t) d\vec{k} + \int_\infty C_-(\vec{k}) \exp(i\vec{k} \cdot \vec{r} + iv_F k t) d\vec{k}, \qquad (5)$$

$$\Psi_2(\vec{r},t) = \int_\infty C_+(\vec{k})(k_x \pm ik_y)k^{-1} \exp(i\vec{k} \cdot \vec{r} - iv_F k t) d\vec{k}$$
$$- \int_\infty C_-(\vec{k})(k_x \pm ik_y)k^{-1} \exp(i\vec{k} \cdot \vec{r} + iv_F k t) d\vec{k}, \qquad (6)$$

where + and – in $\pm$ are respectively for $K$ and $K'$, and $C_+(\vec{k})$ and $C_-(\vec{k})$ are

the superposition coefficients. Accordingly, $|\Psi_1|^2$ and $|\Psi_2|^2$ respectively represent electronic probability density of the wave-packet state at $A$-type atoms and $B$-type atoms. The summation

$$\rho(\vec{r},t) = |\Psi_1|^2 + |\Psi_2|^2 \tag{7}$$

gives the average electronic probability density, with details related to the lattice structure being neglected.

Wave-packet states composed of $\Psi_1(\vec{r},t)$ and $\Psi_2(\vec{r},t)$ satisfy the time-dependent Dirac equation. Hence by working out integrals in Eqs. (5) and (6) for different $t$, one obtains exactly the evolution of the wave packets. The direct superposition of eigen-states is convenient for study in wave-vector space. One can choose different states to compose the wave packets. In general, the wave-packet evolution is obtained by solving the time-dependent Dirac equation for a given initial state, through Green's function method[8] or the time-evolution operator.[10] In that case the role of initial states is easily investigated. The initial state of a wave packet can be obtained directly from Eqs. (5) and (6) by setting $t=0$. The superposition coefficients, on the other hand, can be obtained from the initial state by a Fourier transform

$$C_+(\vec{k}) = \frac{1}{2(2\pi)^3} \int_\infty [\Psi_1(\vec{r},0) + (k_x \mp ik_y)k^{-1}\Psi_2(\vec{r},0)]\exp(-i\vec{k}\cdot\vec{r})d\vec{r}, \tag{8}$$

$$C_-(\vec{k}) = \frac{1}{2(2\pi)^3} \int_\infty [\Psi_1(\vec{r},0) - (k_x \mp ik_y)k^{-1}\Psi_2(\vec{r},0)]\exp(-i\vec{k}\cdot\vec{r})d\vec{r}, \tag{9}$$

where – and + in $\mp$ are respectively for $K$ and $K'$.

We use polar coordinates $(k,\theta)$ and $(r,\varphi)$ with $k_x = k\cos\theta$, $k_y = k\sin\theta$, $x = r\cos\varphi$, and $y = r\sin\varphi$ to calculate integrals in Eqs. (5) and (6). If $C_+ = C_+(k)$ and $C_- = C_-(k)$ are independent of $\theta$, then both $|\Psi_1|$ and $|\Psi_2|$ are independent of $\varphi$, and both $\Psi_1$ and $\Psi_2$ are identical for $K$ and $K'$. Hence the probability density $\rho = \rho(r,t)$ exhibits circular symmetry and demonstrates no difference for the two kinds of Dirac points. The proof is presented in Appendix A. Besides, if one takes coefficients $C_+(k)\exp(\mp i\theta)$ and $-C_-(k)\exp(\mp i\theta)$, then $\Psi_1$ and $\Psi_2$ exchange one another. The two sublattices are thus equivalent for the wave-packet states.

## III. CALCULATION RESULTS

First we superpose the positive states only and take the coefficients

$$C_+(k,\theta) = N\exp[-k^2/(\Delta k)^2], \qquad C_-(k,\theta) = 0, \qquad (10)$$

where $N$ is the normalization constant and $\Delta k > 0$ is a small quantity. Calculation results are presented in Fig. 1(a). The probability density has circular symmetry. At $t = 0$, the probability density $\rho(r,0)$ is a Gaussian-like function located at $r = 0$. As time increases, the central peak decreases monotonously to zero and an annular peak appears nearby. For the annular peak, its distance to the center increases with time; its height decreases with time; and its width remains unchanged. The wave packet evolves form the initial Gaussian form and propagates to all directions like an expanding ripple-ring on water surface. For considerably large $t$, the speed of the ripple-ring is fixed at $v_F$.

Next we superpose negative states only and take the coefficients

$$C_+(k,\theta) = 0, \qquad C_-(k,\theta) = N\exp[-k^2/(\Delta k)^2]. \qquad (11)$$

Calculation results demonstrate that the probability density is the same as that in the superposition of positive states only. It can be proven that the superposition of positive states only and the superposition of negative states only lead to exactly the same probability density. The proof is presented in Appendix B.

Thirdly we superpose both positive and negative states and take the coefficients

$$C_+(k,\theta) = C_-(k,\theta) = \frac{N}{2}\exp[-k^2/(\Delta k)^2], \tag{12}$$

With constant $N = 1/\pi\sqrt{2\pi}\Delta k$, the normalized initial envelope functions are $\Psi_1(\vec{r},0) = (\Delta k/\sqrt{2\pi})\exp[-(\Delta k)^2 r^2/4]$ and $\Psi_2(\vec{r},0) = 0$, and the probability density is a Gaussian function located at $r = 0$ and with width[17] $\sigma = 1/\Delta k$

$$\rho(r,0) = \frac{(\Delta k)^2}{2\pi}\exp[-(\Delta k)^2 r^2/2]. \tag{13}$$

Calculation results are presented in Fig. 1(b). As time increases, the central peak decreases and two annular peaks successively appear nearby. When the first peak is forming, the central probability density decreases monotonously to zero. Then the second peak starts to form. The central probability density first increases from zero to a maximum and then decreases to zero again. For the two annular peaks, their distances to the center increase with time; their heights decrease with time; and their interval and widths remain unchanged. For considerably large $t$, both peaks have the fixed speed $v_F$. The probability density has circular symmetry. This is the picture of two ripple-rings that expand to all directions with the same speed.

During the formation of the first peak, the probability transfers from the central peak to the first ripple-ring. During the formation of the second peak, it first transfers from the first ripple-ring back to the central peak, and then from the central peak to the second ripple-ring. Hence probability alternates between the central peak and the ripple-rings. The central probability density undergoes a transient oscillation. The formation of ripple-rings and the probability-density oscillation are demonstrated in Fig. 2.

Graphene has C-C bond length $d = 0.142\ nm$, lattice constant $a = 0.246\ nm$, and the length of basic reciprocal vectors $b = 29.499\ nm^{-1}$. For the superposition of both positive and negative states, we choose $\Delta k = 0.42 \, nm^{-1}$ so that the wave packet covers dozens of lattice cells. For the superposition of positive states or negative states only, we choose $\Delta k = 0.59 \, nm^{-1}$ so that the initial wave packet is approximately the same as that for the superposition of both positive and negative states. Other values of $\Delta k$ are also attempted. As in usual crystals, a larger $\Delta k$ leads to a narrower wave packet in real space and a smaller $\Delta k$ to a wider one, which is common to all wave packets.

The axial symmetry of the energy dispersion leads to the circular symmetry of the wave packets in real space. The proof in Appendix A that the probability density is independent of $\varphi$ is based on the prerequisite that the energy in Eq. (2) is independent of $\theta$. The fixed speed and widths of the annular peaks, however, are attributed to the linearity of the energy dispersion. Calculations indicate that if this linearity is destroyed, both the speed and the widths of the annular peaks are no longer invariant.

The superposition of positive states only and the superposition of negative states only lead to exactly the same wave packets, although a positive state and its negative counterpart have opposite velocities. The reason is that in the superposition, wave vectors constitute a region containing a Dirac point at the center. In this case, effects of the superposition cannot be ascertained directly form the characteristics of every single state, as the Dirac point is a singularity of the energy dispersion.

Because the energy dispersion has two branches, the positive and negative states interfere in the superposition. As a result, the probability density is not the simple summation of those for positive or negative states only. Instead, multiple ripple-rings form and transient oscillation occurs at the center. If only positive or negative states are superposed, one annular peak remains in the propagation. If both positive and negative states are superposed, in general two annular peaks remain.

During the initial propagation, the two envelope functions $\Psi_1$ and $\Psi_2$ may contribute differently to the probability density, depending on the superposition coefficients. Besides, both $\int_\infty |\Psi_1|^2 d\vec{r}$ and $\int_\infty |\Psi_2|^2 d\vec{r}$ vary with time. The summation $\int_\infty |\Psi_1|^2 d\vec{r} + \int_\infty |\Psi_2|^2 d\vec{r}$, however, is a constant independent of time. As time increase, $|\Psi_1|$ and $|\Psi_2|$ asymptotically become identical. Hence eventually electrons are evenly distributed at two sublattices, regardless of the superposition coefficients. We note that in a crystal, wave-packets can be constructed by $\psi = \int_\infty C(\vec{k}) \exp(i\vec{k}\cdot\vec{r} - iEt/\hbar) d\vec{k}$. For graphene, two envelope functions must be included because the energy dispersion (2) is based on two sublattices. The decrease of the peak-heights in the propagation is the result of the probability normalization.

Different superposition coefficients lead to versatile wave-packet states. For instance, the step-like coefficients

$$C_+(k,\theta) = C_-(k,\theta) = \begin{cases} N & k \leq \Delta k \\ 0 & k > \Delta k \end{cases}, \qquad (14)$$

lead to more ripple-rings for the same initial Gaussian-like wave packet. The alternation of probability between the central peak and the ripple-rings is more complicated, and the central probability density undergoes more sustained oscillation [Fig. 2(b) inset]. Coefficients that depend on $\theta$ lead to annular wave packets that have no exact circular symmetry. However, the main features of the wave packets are the same.

## IV. ANNULAR WAVE PACKETS IN A MAGNETIC FIELD

In general researches of wave packets in a magnetic field, the energy is expanded into a Taylor series of the quantum number. The quasi-classical oscillation period and the revival period are obtained from the expansion coefficients.[10,12,18] For graphene, however, according to Landau energies $E_\pm = \pm v_F \sqrt{2n\hbar eB}$ where $B$ is the magnetic field strength and $n = 0, 1, 2, \cdots$ is the quantum number, derivatives $dE/dn$, $d^2E/dn^2$, etc. do not exist at $n = 0$. Hence one cannot expand the energy into a Taylor series if $n = 0$ is the central quantum number. This is the case for our wave packets and we therefore investigate their propagation by direct superposition.

For graphene in a perpendicular magnetic field with vector potential $\vec{A}$, the two envelope functions $\psi_1$ and $\psi_2$ of the electronic eigen-states satisfy

$$v_F \begin{pmatrix} 0 & \hat{p}_x + eA_x \mp i(\hat{p}_y + eA_y) \\ \hat{p}_x + eA_x \pm i(\hat{p}_y + eA_y) & 0 \end{pmatrix} \begin{pmatrix} \psi_1 \\ \psi_2 \end{pmatrix} = E \begin{pmatrix} \psi_1 \\ \psi_2 \end{pmatrix}, \quad (15)$$

where the upper and lower signs in $\mp$ and $\pm$ are respectively for $K$ and $K'$.[1] For circular symmetry we use the polar coordinates $(r, \varphi)$ and adopt the symmetric gauge for the vector potential $A_x = -(Br/2)\sin\varphi$, $A_y = (Br/2)\cos\varphi$. After some manipulation Eq. (15) becomes

$$v_F \left( -i\hbar \frac{\partial \psi_2}{\partial r} \mp \frac{\hbar}{r} \frac{\partial \psi_2}{\partial \varphi} \mp \frac{ieB}{2} r \psi_2 \right) \times \exp(\mp i\varphi) = E\psi_1, \quad (16)$$

$$v_F \left( -i\hbar \frac{\partial \psi_1}{\partial r} \pm \frac{\hbar}{r} \frac{\partial \psi_1}{\partial \varphi} \pm \frac{ieB}{2} r \psi_1 \right) \times \exp(\pm i\varphi) = E\psi_2. \quad (17)$$

Hence $\psi_1$ and $\psi_2$ respectively satisfy

$$-\hbar^2 \nabla^2 \psi_1 + \frac{e^2 B^2}{4} r^2 \psi_1 - i\hbar eB \frac{\partial \psi_1}{\partial \varphi} = \left( \frac{E^2}{v_F^2} \mp \hbar eB \right) \psi_1, \quad (18)$$

$$-\hbar^2 \nabla^2 \psi_2 + \frac{e^2 B^2}{4} r^2 \psi_2 - i\hbar eB \frac{\partial \psi_2}{\partial \varphi} = \left( \frac{E^2}{v_F^2} \pm \hbar eB \right) \psi_2, \quad (19)$$

where $\nabla^2 = \partial^2/\partial r^2 + (1/r)\partial/\partial r + (1/r^2)\partial^2/\partial \varphi^2$. The eigen-functions of Eqs. (18) and (19) are the products of those for 2D harmonic Hamiltonian and those for $z$-component of angular momentum.[19,20] For circular symmetry we consider the simplest case where one envelope function has zero $z$-component of angular momentum. For $K$ one obtains the orthonormal eigen-functions

$$\psi_1^\pm(r,\varphi,n) = \pm \frac{i\sqrt{n}}{2\sqrt{2\pi}L^2} r \exp(-r^2/4L^2) \times F(-n+1,2,r^2/2L^2) \times \exp(-i\varphi), \quad (20)$$

$$\psi_2^\pm(r,n) = \frac{1}{2\sqrt{\pi}L} \exp(-r^2/4L^2) \times F(-n,1,r^2/2L^2), \quad (21)$$

and the eigen-energies

$$E_\pm(n) = \pm \frac{\sqrt{2n}\hbar v_F}{L}. \quad (22)$$

Here, $n = 0,1,2,\cdots$ is the quantum number; $L = \sqrt{\hbar/eB}$ is the magnetic length; $F(\alpha,\gamma,z)$ is the confluent hypergeometric function $F(\alpha,\gamma,z) = 1 + \sum_{j=1}^{+\infty} (j!)^{-1}\alpha(\alpha+1)\cdots(\alpha+j-1)\gamma^{-1}(\gamma+1)^{-1}\cdots(\gamma+j-1)^{-1}z^j$ which becomes polynomials in our case; and + and – in $\pm$ are respectively for positive and negative eigen-energies. The envelope functions of wave-packet states are the superposition of $\psi_1^\pm$ and $\psi_2^\pm$

$$\Psi_1(\vec{r},t) = \sum_{n=0}^{+\infty} C_+(n)\psi_1^+(r,\varphi,n)\exp[-iE_+(n)t/\hbar]$$
$$+ \sum_{n=0}^{+\infty} C_-(n)\psi_1^-(r,\varphi,n)\exp[-iE_-(n)t/\hbar], \quad (23)$$

$$\Psi_2(\vec{r},t) = \sum_{n=0}^{+\infty} C_+(n)\psi_2^+(r,n)\exp[-iE_+(n)t/\hbar]$$
$$+ \sum_{n=0}^{+\infty} C_-(n)\psi_2^-(r,n)\exp[-iE_-(n)t/\hbar], \quad (24)$$

where $C_+(n)$ and $C_-(n)$ are superposition coefficients satisfying $\sum_{n=0}^{+\infty}[|C_+(n)|^2 + |C_-(n)|^2] = 1$. The wave-packet states satisfy the time-dependent Dirac equation. Hence by working out summations in Eqs. (23) and (24) for different $t$, one obtains exactly the evolution of the wave packet in the magnetic field.

The initial state of a wave packet can be obtained directly from Eqs. (23) and (24) by setting $t = 0$. The superposition coefficients, on the other hand, can be obtained according to the orthonormality of $\psi_1^\pm$ and $\psi_2^\pm$. They are

$$C_+(n) = -\int_\infty \Psi_1(\vec{r},0)\psi_1^+(r,\varphi,n)\exp(2i\varphi)d\vec{r} + \int_\infty \Psi_2(\vec{r},0)\psi_2^+(r,n)d\vec{r}, \tag{25}$$

$$C_-(n) = -\int_\infty \Psi_1(\vec{r},0)\psi_1^-(r,\varphi,n)\exp(2i\varphi)d\vec{r} + \int_\infty \Psi_2(\vec{r},0)\psi_2^-(r,n)d\vec{r}. \tag{26}$$

We consider the initial functions $\Psi_1(\vec{r},0) = 0$ and $\Psi_2(\vec{r},0) = (1/\sqrt{2\pi}\sigma)\exp(-r^2/4\sigma^2)$ that lead to the initial probability density

$$\rho(r,0) = \frac{1}{2\pi\sigma^2}\exp(-r^2/2\sigma^2) \tag{27}$$

with width $\sigma$. The superposition coefficients are real numbers

$$C_+(n) = C_-(n) = \frac{\sqrt{2\pi}}{\sigma}\int_0^{+\infty} r\exp(-r^2/4\sigma^2)\psi_2^\pm(r,n)dr. \tag{28}$$

Calculations indicate that $|C_+(n)|=|C_-(n)|$ decreases monotonously as $n$ increases, with $|C_+(0)|=|C_-(0)|$ having the largest value. Hence the wave packet has the central quantum number $n = 0$.

Both the width of the initial Gaussian function $\sigma$ and the strength of magnetic field $B$ affect the wave-packet evolution. For different values of the ratio

$$\beta = \frac{\sigma}{L}, \tag{29}$$

the evolution demonstrates different features. For $\beta = 1$, the wave packet does not vary with time, because the initial state happens to be the eigen-state with $n = 0$. For $\beta \sim 1$, the width and height of the central peak oscillate periodically, and the wave packet does not propagate. This is demonstrated in Fig. 3(a). For $\beta \ll 1$, the evolution is more complicated. Starting from the initial Gaussian form, at first the central peak decreases and expanding ripple-rings form. Nevertheless, the ripple-rings do not expand all the time. They later disappear and other ripple-rings that shrink to the center form. As time increases, expanding and shrinking ripple-rings form and disappear alternatively, and their radii and heights oscillate with time. At times the wave packet resumes a Gaussian form at the center, only the Gaussian form is wider than the initial one. This gives a picture of breathing ripple-rings. The spread of the wave packet does not increase unlimitedly, and probability alternates between the central peak and the ripple-rings persistently. Different from the case of free graphene, the wave packet is confined by the magnetic field. For $\beta \gg 1$, the central peak first increases to a maximum and then decreases. After that, expanding and shrinking ripple-rings form and disappear alternatively in a limited spread. At times the wave packet resumes the Gaussian form that is narrower than the initial one. The probability-density oscillation is demonstrated in Fig. 3, and complicated wave-packet evolution is demonstrated in Fig. 4.

For $\beta \sim 1$, the coefficient $|C_+(n)|=|C_-(n)|$ decreases rapidly as $n$ increases. In Eqs. (23) and (24), terms with smaller $n$ contribute much greater to the probability density than terms with greater $n$. If we preserve first two terms in the probability density, one has $\rho(r,t) \approx 4C_+^2(0)[\psi_2^+(r,0)]^2 + 8C_+(0)C_+(1)\psi_2^+(r,0)\psi_2^+(r,1)\cos[E_+(1)t/\hbar]$. Hence the probability density mainly has a constant part and a periodical part. The period is

$$T_1 = \frac{2\pi\hbar}{E_+(1)} = \frac{\sqrt{2}\pi L}{v_F}, \qquad (30)$$

which is verified in Fig. 3(a). For $\beta \ll 1$ or $\beta \gg 1$, the coefficient $|C_+(n)|=|C_-(n)|$ decreases slowly as $n$ increases, and effects of different coefficients cannot be separated distinctly. In Eqs. (23) and (24), many terms with different period $T_n = 2\pi\hbar/E_+(n) = \sqrt{2}\pi n^{-1/2} L v_F^{-1}$ contribute to the probability density considerably. As $T_n \propto n^{-1/2}$, the frequencies cannot be multiples of a minimum one. As a result, the probability density varies without exact periodicity. The spread of the wave packet is mainly determined by the exponential function $\exp(-r^2/4L^2)$ in $\psi_1^\pm$ and $\psi_2^\pm$. Hence a strong magnetic field leads to a narrow spread. As magnetic field is usually weak, we take a wider Gaussian function with $\sigma = 24\,nm$ for the calculations. For this $\sigma$, $\beta = 1$ corresponds to $B = 1.15\,T$.

For $K'$ the orthonormal eigen-functions are obtained by exchanging $\psi_1^\pm$ and $\psi_2^\pm$ in Eqs. (20) and (21), and the eigen-energies are still expressed in Eq. (22). The two envelope functions $\Psi_1(\vec{r},t)$ and $\Psi_2(\vec{r},t)$ thus exchange one another and the wave packet remains the same. Due to the two kinds of Dirac points the two sublattices demonstrate no discrimination to the wave-packet states.

Magnetic field provides a means to prepare annular wave-packet states as its eigen-states. One may then change the field and study the wave-packet evolution. In general, wave packets are not stationary states and are expected to have electromagnetic radiation. For the localized wave packets, their emission of electromagnetic waves due to the trembling motion is investigated in terms of their dipole-moment variation.[9,15] Hence the essential factor is the electronic probability variation. For the annular wave packets, although their average coordinates do not vary with time, they still demonstrate probability oscillation, and thus may have electromagnetic radiation. One may expect an axial electrical field with axis perpendicular to the graphene plane and passing through the center, and an annular magnetic field whose field lines are concentric circles parallel to the ripple-rings. In an external magnetic field, the field of the wave packets may be more complicated, as circular current exists. The electromagnetic radiation may be studied according to the classical retarded potentials, similar to the approach in Refs. 9 and 15. Annular wave packets of graphene in a magnetic field provide us with trapped electronic states. Like Rydberg states in atoms, such states may be useful for research on the border of quantum and classical realms.[21] When probability alternates between the central peak and ripple-rings, one may expect similar effects as those for the trembling motion of localized wave packets. The foundation of the effects, however, is the features of the annular wave packets.

## V. CONCLUSIONS

In conclusion, wave packets in graphene at Dirac points are investigated. Although their wave vectors are confined to a localized region in wave-vector space, the wave packets are quasi-extended states in real space, and probability is distributed in propagating ripple-rings. In a magnetic field, the annular wave packets demonstrate more complicated evolution in a confined region. Features of the wave packets directly reflect characteristics of the electronic structure of graphene, especially the energy dispersion near the Dirac points, and Landau energies and wave functions near $n=0$. The features constitute the foundation for further investigation on possible physical effects.

## APPENDIX A

With polar coordinates, Eqs. (5) and (6) become

$$\Psi_1(r,\varphi,t) = \int_0^{2\pi} d\theta \int_0^{+\infty} C_+(k,\theta) k \exp[ikr\cos(\theta-\varphi) - iv_F kt] dk \\ + \int_0^{2\pi} d\theta \int_0^{+\infty} C_-(k,\theta) k \exp[ikr\cos(\theta-\varphi) + iv_F kt] dk, \quad (A1)$$

$$\Psi_2(r,\varphi,t) = \int_0^{2\pi} d\theta \int_0^{+\infty} C_+(k,\theta) k \exp[ikr\cos(\theta-\varphi) - iv_F kt \pm i\theta] dk \\ - \int_0^{2\pi} d\theta \int_0^{+\infty} C_-(k,\theta) k \exp[ikr\cos(\theta-\varphi) + iv_F kt \pm i\theta] dk. \quad (A2)$$

where + and - in $\pm$ are respectively for $K$ and $K'$. Suppose $C_+(k,\theta) = C_+(k)$ and $C_-(k,\theta) = C_-(k)$ are independent of $\theta$. By applying transform $\theta_1 = \theta - \varphi$ one obtains

$$\Psi_1 = \int_{-\varphi}^{2\pi-\varphi} d\theta_1 \int_0^{+\infty} k C_+(k) \exp[ikr\cos\theta_1 - iv_F kt] dk \\ + \int_{-\varphi}^{2\pi-\varphi} d\theta_1 \int_0^{+\infty} k C_-(k) \exp[ikr\cos\theta_1 + iv_F kt] dk, \quad (A3)$$

$$\Psi_2 = \exp(\mp i\varphi)\int_{-\varphi}^{2\pi-\varphi}d\theta_1\int_0^{+\infty}kC_+(k)\exp[ikr\cos\theta_1 - iv_F kt \pm i\theta_1]dk$$
$$-\exp(\mp i\varphi)\int_{-\varphi}^{2\pi-\varphi}d\theta_1\int_0^{+\infty}kC_-(k)\exp[ikr\cos\theta_1 + iv_F kt \pm i\theta_1]dk.$$
(A4)

The integrands in Eqs. (A3) and (A4) are periodical functions of $\theta_1$ with the period being $2\pi$. One has $\int_{-\varphi}^{2\pi-\varphi}d\theta_1\cdots = \int_0^{2\pi}d\theta_1\cdots$ and

$$\Psi_1 = \int_0^{2\pi}d\theta_1\int_0^{+\infty}kC_+(k)\exp[ikr\cos\theta_1 - iv_F kt]dk$$
$$+\int_0^{2\pi}d\theta_1\int_0^{+\infty}kC_-(k)\exp[ikr\cos\theta_1 + iv_F kt]dk,$$
(A5)

$$\Psi_2 = \exp(\mp i\varphi)\int_0^{2\pi}d\theta_1\int_0^{+\infty}kC_+(k)\exp[ikr\cos\theta_1 - iv_F kt \pm i\theta_1]dk$$
$$-\exp(\mp i\varphi)\int_0^{2\pi}d\theta_1\int_0^{+\infty}kC_-(k)\exp[ikr\cos\theta_1 + iv_F kt \pm i\theta_1]dk.$$
(A6)

Hence $|\Psi_1(\vec{r},t)|$ and $|\Psi_2(\vec{r},t)|$ are independent of $\varphi$.

It is obvious that $\Psi_1$ is the same for $K$ and $K'$. In Eq. (A6), integrals with respect to $\theta_1$ are $\int_0^{2\pi}\cos(kr\cos\theta_1 \pm \theta_1)d\theta_1$ and $\int_0^{2\pi}\sin(kr\cos\theta_1 \pm \theta_1)d\theta_1$, where + and - in $\pm$ are respectively for $K$ and $K'$. By applying the transform $\theta_2 = -\theta_1$ one has $\int_0^{2\pi}\cos(kr\cos\theta_1 - \theta_1)d\theta_1 = \int_{-2\pi}^{0}\cos(kr\cos\theta_2 + \theta_2)d\theta_2$. For the periodicity of $\cos(kr\cos\theta_2 + \theta_2)$ one has $\int_0^{2\pi}\cos(kr\cos\theta_1 - \theta_1)d\theta_1 = \int_0^{2\pi}\cos(kr\cos\theta_2 + \theta_2)d\theta_2$. Similarly one has $\int_0^{2\pi}\sin(kr\cos\theta_1 - \theta_1)d\theta_1 = \int_0^{2\pi}\sin(kr\cos\theta_2 + \theta_2)d\theta_2$. Hence $\Psi_2$ is the same for $K$ and $K'$.

## APPENDIX B

In the superposition of positive states only, by supposing $C_+(k,\theta) = C(k)$, $C_-(k,\theta) = 0$, and $\varphi = 0$ for Eqs. (A1) and (A2) one has

$$\Psi_1^+ = \int_0^{2\pi}d\theta\int_0^{+\infty}C(k)k\exp(ikr\cos\theta - iv_F kt)dk,$$
(B1)

$$\Psi_2^+ = \int_0^{2\pi} d\theta \int_0^{+\infty} C(k)k \exp(ikr\cos\theta - iv_F kt \pm i\theta)dk .\tag{B2}$$

Or equivalently,

$$\Psi_1^+ = \int_0^{2\pi} d\theta \int_0^{+\infty} C(k)k \exp(-iv_F kt)[\cos(kr\cos\theta) + i\sin(kr\cos\theta)]dk ,\tag{B3}$$

$$\Psi_2^+ = \int_0^{2\pi} d\theta \int_0^{+\infty} C(k)k \exp(-iv_F kt)[\cos(kr\cos\theta \pm \theta) + i\sin(kr\cos\theta \pm \theta)]dk .\tag{B4}$$

Because $\int_0^{2\pi} \sin(kr\cos\theta)d\theta = \int_0^{\pi/2} + \int_{\pi/2}^{\pi} + \int_{\pi}^{3\pi/2} + \int_{3\pi/2}^{2\pi}$ , by respectively applying transforms $\theta = \pi - \theta_2$, $\theta = \pi + \theta_3$, and $\theta = 2\pi - \theta_4$ to $\int_{\pi/2}^{\pi}$, $\int_{\pi}^{3\pi/2}$, and $\int_{3\pi/2}^{2\pi}$, one obtains $\int_0^{2\pi} \sin(kr\cos\theta)d\theta = \int_0^{\pi/2} - \int_0^{\pi/2} - \int_0^{\pi/2} + \int_0^{\pi/2} = 0$. By a similar way one can prove $\int_0^{2\pi} \cos(kr\cos\theta \pm \theta)d\theta = 0$. Hence Eqs. (B3) and (B4) become

$$\Psi_1^+ = \int_0^{2\pi} d\theta \int_0^{+\infty} C(k)k \exp(-iv_F kt)\cos(kr\cos\theta)dk ,\tag{B5}$$

$$\Psi_2^+ = i\int_0^{2\pi} d\theta \int_0^{+\infty} C(k)k \exp(-iv_F kt)\sin(kr\cos\theta \pm \theta)dk .\tag{B6}$$

In the superposition of negative states only, by supposing $C_+(k,\theta) = 0$, $C_-(k,\theta) = C(k)$, and $\varphi = 0$, by a similar procedure one has

$$\Psi_1^- = \int_0^{2\pi} d\theta \int_0^{+\infty} C(k)k \exp(iv_F kt)\cos(kr\cos\theta)dk ,\tag{B7}$$

$$\Psi_2^- = -i\int_0^{2\pi} d\theta \int_0^{+\infty} C(k)k \exp(iv_F kt)\sin(kr\cos\theta \pm \theta)dk .\tag{B8}$$

From Eqs. (B5), (B6), (B7), and (B8), one has $\Psi_1^- = (\Psi_1^+)^*$ and $\Psi_2^- = (\Psi_2^+)^*$. Hence $|\Psi_1^-(r,t)| = |\Psi_1^+(r,t)|$, $|\Psi_2^-(r,t)| = |\Psi_2^+(r,t)|$, and the probability densities are the same.

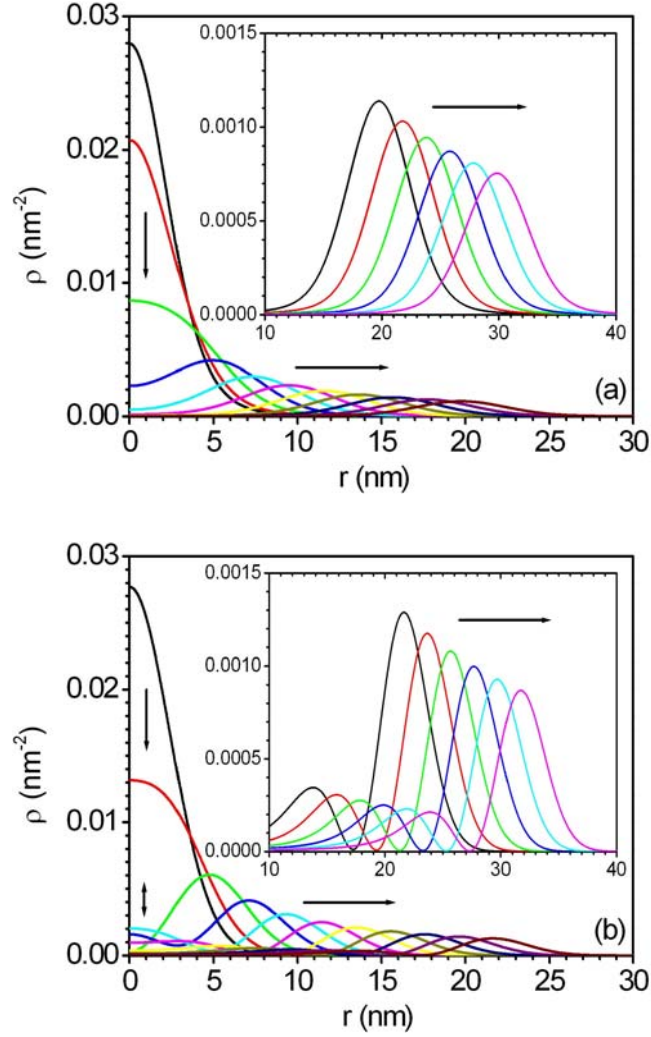

Fig. 1. Probability density as a function of the radium, demonstrating the formation of ripple-rings and their propagation in free graphene. Curves indicate wave packet at time $t = 0$, $t = 2\ fs$, $t = 4\ fs$, $t = 6\ fs$, $t = 8\ fs$, $t = 10\ fs$, $t = 12\ fs$, $t = 14\ fs$, $t = 16\ fs$, $t = 18\ fs$, and $t = 20\ fs$ with step $\Delta t = 2\ fs$. Curves in Insets indicate wave packet at time $t = 20\ fs$, $t = 22\ fs$, $t = 24\ fs$, $t = 26\ fs$, $t = 28\ fs$, and $t = 30\ fs$ with step $\Delta t = 2\ fs$. (a) For the wave packet formed from the positive states only according to the superposition coefficients in Eq. (10). (b) For the wave packet formed from both positive and negative states according to the superposition coefficients in Eq. (12).

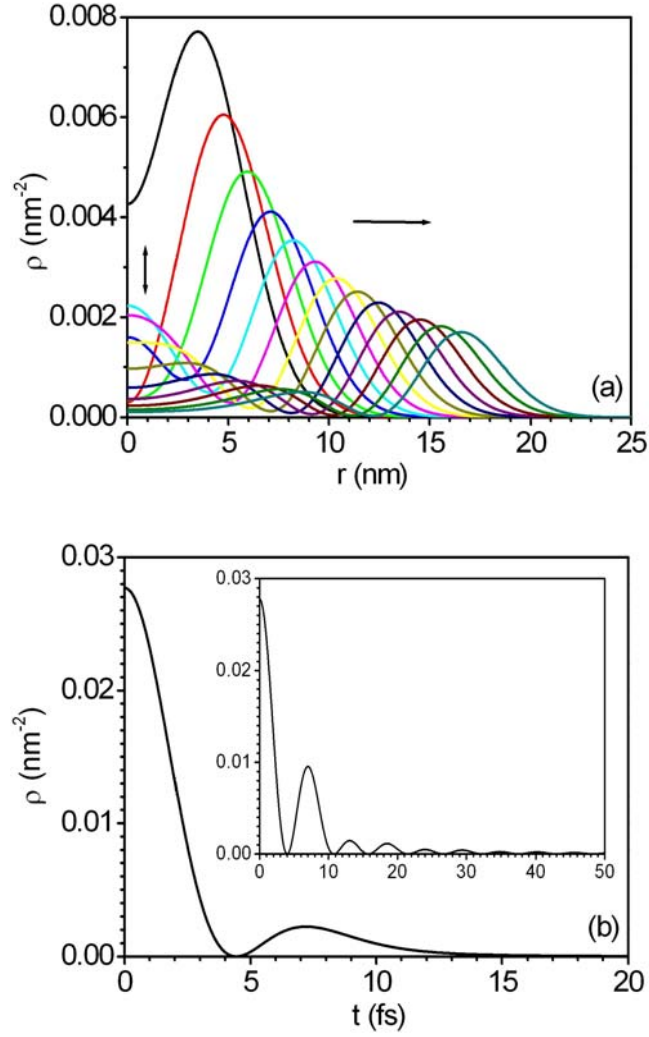

Fig. 2. (a) Probability density as a function of the radium for the wave packet formed form both positive and negative states according to the superposition coefficients in Eq. (12) in free graphene, at time $t = 3 \sim 15\ fs$ with step $\Delta t = 1\ fs$. It demonstrates the formation of the second ripple-ring. (b) Probability density at the center $r = 0$ as a function of time for the same wave packet, demonstrating the probability-density oscillation. Inset of (b): For the wave packet formed according to the superposition coefficients in Eq. (14).

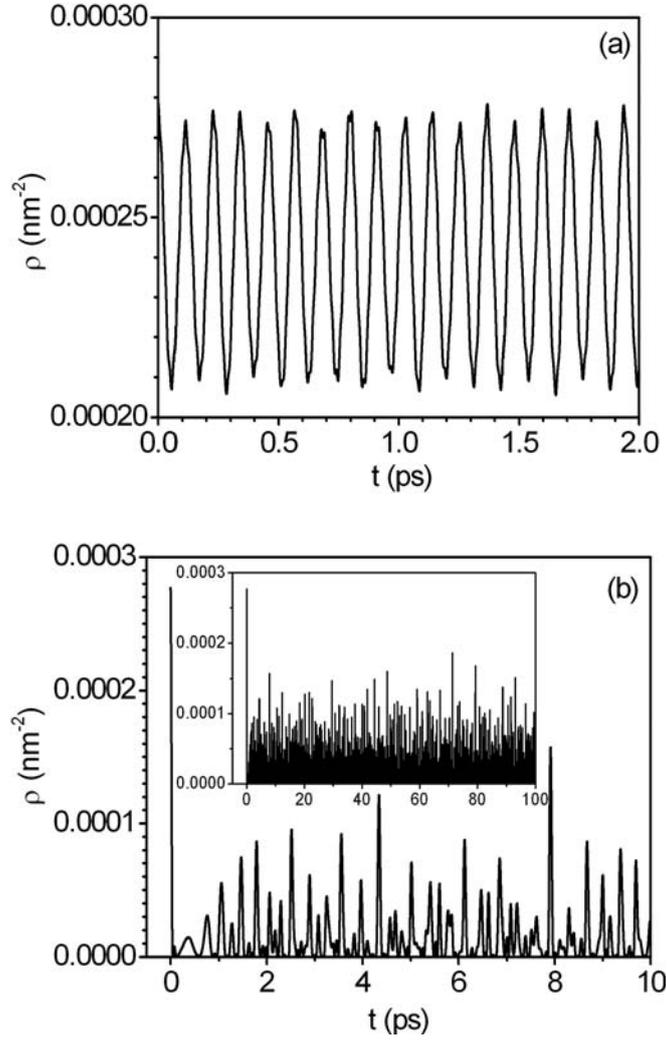

Fig. 3. Probability density at the center $r=0$ as a function of time, for graphene in a magnetic field. (a) For the wave packet with $\sigma=24\,nm$, $B=1T$, and $\sigma/L=0.93$. It demonstrates the periodical oscillation of the central peak and the period is determined by Eq. (30) as $T_1=0.114\,ps$. (b) For the wave packet with $\sigma=24\,nm$, $B=0.1T$, and $\sigma/L=0.3$. It demonstrates more complicated oscillation of the probability density.

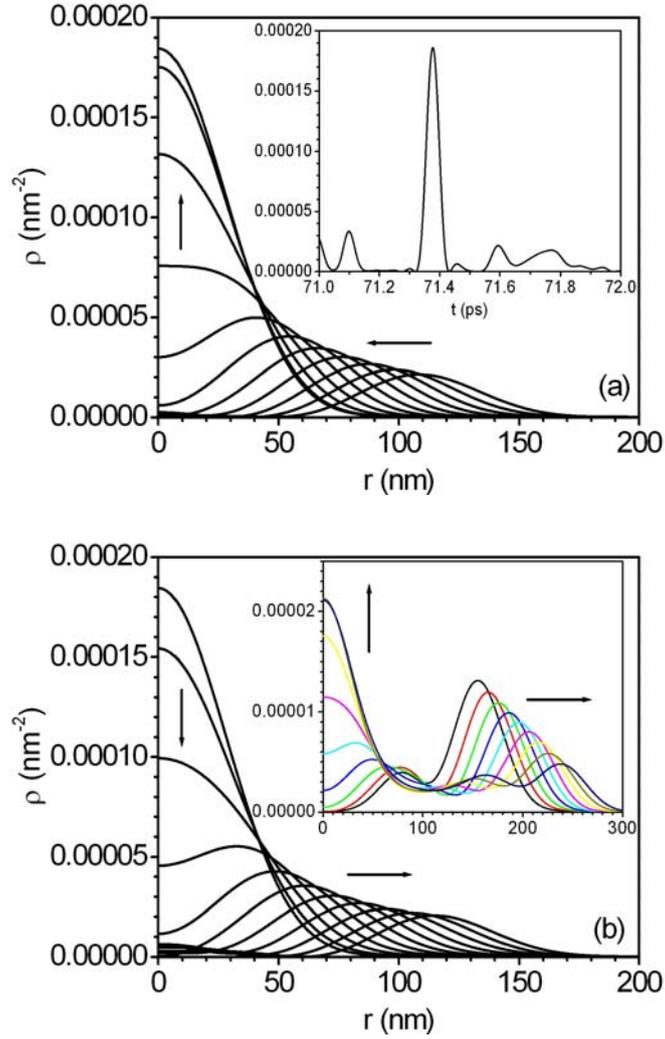

Fig. 4. Probability density as a function of the radium, for graphene in a magnetic field with $B = 0.1\,T$. The wave packet has $\sigma = 24\,nm$ and $\sigma/L = 0.3$. (a) Wave packet at time $t = 71280\,fs$, $t = 71290\,fs$, $t = 71300\,fs$, $t = 71310\,fs$, $t = 71320\,fs$, $t = 71330\,fs$, $t = 71340\,fs$, $t = 71350\,fs$, $t = 71360\,fs$, $t = 71370\,fs$, and $t = 71380\,fs$ with step $\Delta t = 10\,fs$, demonstrating shrinking ripple-rings. (b) Wave packet at time $t = 71380\,fs$, $t = 71390\,fs$, $t = 71400\,fs$, $t = 71410\,fs$, $t = 71420\,fs$, $t = 71430\,fs$, $t = 71440\,fs$, $t = 71450\,fs$, $t = 71460\,fs$, $t = 71470\,fs$, and $t = 71480\,fs$ with step $\Delta t = 10\,fs$, demonstrating expanding ripple-rings. The wave packet resumes the Gaussian form at $t = 71380\,fs$.

Inset of (a): Probability density at the center $r = 0$ as a function of time from $t = 71000 \ fs$ to $t = 72000 \ fs$. The peak corresponds to the recovery of the wave packet. Inset of (b): Wave packet at time $t = 71520 \ fs$, $t = 71530 \ fs$, $t = 71540 \ fs$, $t = 71550 \ fs$, $t = 71560 \ fs$, $t = 71570 \ fs$, $t = 71580 \ fs$, $t = 71590 \ fs$, and $t = 71600 \ fs$ with step $\Delta t = 10 \ fs$, demonstrating probability transfer from the ripple-ring to the central peak.